\documentclass[twocolumn,showpacs,preprintnumbers,amsmathamssymb]{revtex4}

\usepackage{graphicx}
\usepackage{epsfig}

\begin{document}

\title{Multi-gap superconductivity in a BaFe$_{1.84}$Co$_{0.16}$As$_2$ film from optical measurements at terahertz frequencies}
\author{A. Perucchi$^1$, L. Baldassarre$^1$, S. Lupi$^2$
J. Jiang$^3$, J.D. Weiss$^3$, E.E. Hellstrom$^3$,
S. Lee$^4$, C.W. Bark$^4$, C.B. Eom$^4$, M. Putti$^5$, I. Pallecchi$^5$, C. Marini$^6$, P. Dore$^6$}

\affiliation{$^1$ Sincrotrone Trieste S.C.p.A., S.S. 14 km 163.5, in
Area Science Park, 34012 Basovizza (Trieste), Italy}
\affiliation{$^2$ CNR-IOM and Dipartimento di Fisica, Universit\'{a} 
di Roma Sapienza, P.le A.Moro 2, 00185 Rome, Italy}
\affiliation{$^3$ Applied Superconductivity Center,
National High Magnetic Field Laboratory, Florida State University,
2031 East Paul Dirac Drive, Tallahassee, FL 32310, USA}
\affiliation{$^4$ Department of Materials Science and Engineering,
University of Wisconsin-Madison, Madison, WI 53706, USA}
\affiliation{$^5$ CNR-SPIN and Dipartimento di Fisica, Universit\'{a} di Genova, Via
Dodecaneso 33, 16146 Genova, Italy}
\affiliation{$^6$ CNR-SPIN and Dipartimento di
Fisica, Universit\'{a} di Roma ``La Sapienza", P.le A.Moro 2, 00185
Rome, Italy}

\date{\today}

\begin{abstract}
We measured the THz reflectance properties of a high quality
epitaxial thin film of the Fe-based superconductor
BaFe$_{1.84}$Co$_{0.16}$As$_2$ with T$_c$=22.5 K. The film was grown
by pulsed laser deposition on a DyScO$_3$ substrate with an epitaxial
SrTiO$_3$ intermediate layer. The measured $R_S/R_N$ spectrum, i.e.
the reflectivity ratio between the superconducting and normal state
reflectance, suggests the presence of a
superconducting gap $\Delta_A$ close to 15 cm$^{-1}$.
A detailed data analysis shows
that a two-band, two-gap model is necessary to obtain a
good description of the measured $R_S/R_N$ spectrum. The low-energy
$\Delta_A$ gap results to be well determined
($\Delta_A$=15.5$\pm$0.5 cm$^{-1}$),
while the value of the high-energy gap $\Delta_B$ is more uncertain
($\Delta_B$=55$\pm$7 cm$^{-1}$).
Our results provide evidence of two electronic contributions
to the system conductivity with the presence of two optical gaps
corresponding to 2$\Delta/kT_c$ values close to 2 and 7.

\end{abstract}

\maketitle

\section{Introduction}
Since the early days of the discovery of superconductivity in the
iron-pnictide family \cite{kami08}, analogies and differences with
copper-oxide high-T$_c$ superconductors have been extensively
discussed. In particular, the importance of electron correlation
\cite{qazilbash09} and the degree of itinerancy of the charge
carriers \cite{yang09} is still a matter of debate. Closely
connected to that issue, is the problem of the symmetry of the
superconducting order parameter. Indeed nodal gap symmetries are
more liable to accomodate Cooper pairs in the presence of strong
repulsions, than a fully gapped $s$-wave symmetry. Unfortunately,
the determination of the gap properties of pnictides is further
complicated by the fact that multiple bands are crossing the Fermi
level, thus forming semi-metallic electron and hole pockets. While
both nodal \cite{musch09,Goko09,Salem09,Gordon09} and node-less
\cite{terashima09,Samu09,Williams09,hardy09,Tanatar,Luan10} gaps have
been proposed in literature, there is now a growing experimental
evidence favouring the $s$-wave scenario. This finding is consistent
with unconventional superconductivity mediated by spin-fluctuations,
through the so-called $s_{\pm}$ pairing state \cite{Maz08}. In this
scenario, it is of basic importance a detailed knowledge of
symmetry, number and values of the possible superconducting gaps.

In studying the superconducting Fe-based compounds, the Ò122Ó family
obtained by doping of the parent compound AFe$_2$As$_2$ (A=Ba, Sr,
Ca) is particularly suitable to explore the nature of the
superconducting gaps, as it can be prepared in sizeable single
crystals \cite{Luo08} as well as in high quality epitaxial thin
films \cite{lee09,iida09}. Moreover, compounds of the 122 phase
are chemically and structurally simpler and more isotropic than
those of the Ò1111Ó family (ReFeAsO, Re=rare earth), and can have
a fairly high transition temperature. In particular, T$_c$ can be as
high as 25K in the electron doped BaFe$_{2-x}$Co$_x$As$_2$ compound
\cite{Sefat08}, which is the object of the present work. The
superconducting gaps in this system have been measured by various
techniques, namely angle resolved photoemission spectroscopy (ARPES)
\cite{terashima09}, heat capacity \cite{hardy09}, scanning tunneling
spectroscopy \cite{Yin09,Massee09}, penetration depth probed by
$\mu$SR \cite{Williams09}, point contact spectroscopy \cite{Samu09},
Raman scattering \cite{musch09}, and thermal conductivity
\cite{Tanatar}. The results are widely scattered and even in sharp
contrast with one another in terms of symmetry of the gaps, their
number and their values. Indeed, the gaps have been found to be
either nodal \cite{musch09,Goko09,Salem09,Gordon09} or nodeless
\cite{terashima09,Samu09,Williams09,hardy09,Tanatar,Luan10}, isotropic
\cite{terashima09,Samu09} or anisotropic \cite{musch09,Tanatar}. As
to number and values of the gap(s), both a single gap
\cite{Yin09,Massee09,Samu09} and two gaps
\cite{Williams09,terashima09,hardy09,Luan10} have been reported in the
range from 2$\Delta$/KT$_c$ $\simeq$ 2 to $\simeq$ 7. It is
important to note that caution must be used in comparing results
obtained on different samples since doping \cite{terashima09} and
disorder \cite{musch09,Tanatar,Hashimoto09,Carb10} have been indicated to
be likely responsible for the apparent disagreement of literature
data on gap values. Caution must be used also in comparing results
obtained through different techniques since impurity effects, sample
inhomogeneities, and surface off-stoichiometry can differently
affect bulk and surface sensitive measurements.

In general, infrared spectroscopy is a powerful tool to study the
properties of a conducting system since the optical response of the
free-charge carriers is determined by the frequency-dependent
complex conductivity $\tilde{\sigma}=\sigma_{1}+i\sigma_{2}$. In a
simple one-band system, the standard Drude model (with parameters
plasma frequency $\Omega$ and scattering rate $\gamma$) describes
the complex conductivity $\tilde{\sigma}_{N}$ in the normal (N)
state \cite{Burns}. Here it is worth to recall that the $d.c.$
conductivity $\sigma_0$=$\sigma_1(\omega=0)$ is proportional to
$\Omega^2 / \gamma$. In the superconducting (S) state, the standard
BCS model (Mattis-Bardeen equations \cite{matbar}, with parameters
$\sigma_0$ and superconducting gap $\Delta$) can describe the
complex conductivity $\tilde{\sigma}_{S}$ \cite{Tink}. On this
basis, far-infrared/terahertz measurements can be of particular
importance since a mark of the superconducting gap $\Delta$ can be
observed at $\hbar\omega \sim 2\Delta$ (optical gap) for an
isotropic $s$-wave BCS superconductor. For a bulk sample, in
particular, a maximum at the optical gap is expected in the ratio
$R_S/R_N$, where $R_S$ and $R_N$ are the frequency-dependent
reflectances in the superconducting and normal state, respectively
\cite{Tink}.

Infrared spectroscopy has been widely employed in\\
studying Fe-based
compounds \cite{revIR,Drechsler}. In particular,
infrared/ terahertz
measurements on BaFe$_{2-x}$Co$_x$As$_2$ samples
\cite{vanh09,kim09,gorsh09,wu09,Nakai10,Nakam10,Fisher10} have been recently 
employed to study the frequency dependent conductivity and to 
get information on the superconducting gap(s).
However,
there is still poor consensus on the value and, most importantly, on
the number of gaps resulting from the analysis of THz data, as well
as from the analysis of data from different techniques, as discussed
above. In the present work we investigated the multiple-gap
character of the Co-doped BaFe$_2$As$_2$ system by performing
reflectance measurements at terahertz (THz) frequencies on a high
quality film \cite{lee09}. The results of the previous THz studies
will be reported in the following, together with the results of the
present work.

\section{Experimental results}
Epitaxial Co-doped BaFe$_2$As$_2$ films with high transition temperature were
recently obtained by using novel template engineering \cite{lee09}.
The investigated 350 nm thick \\
BaFe$_{1.84}$Co$_{0.16}$As$_2$ (BFCA) film
was grown \emph{in-situ} by using pulsed laser deposition (PLD) on an (110)
oriented DyScO$_3$ (DSO) single crystal substrate with an intermediate
layer (20 unit cell) of SrTiO$_3$ (STO).  Out-of plane  $\theta$-2$\theta$
X-ray diffraction was used to evaluate the quality and the orientation of
the films: The full-width at half-maximum (FWHM) of the 004 rocking curve
is indeed narrow (0.55$^{\circ}$) and the 001 peaks, visible in the spectra,
indicate that the film is grown with the c-axis normal to the substrate.
For further informations on the film growth and on their epitaxial
quality see Ref.\cite{lee09}.

\begin{figure}
\resizebox{0.5\textwidth}{!}{\includegraphics{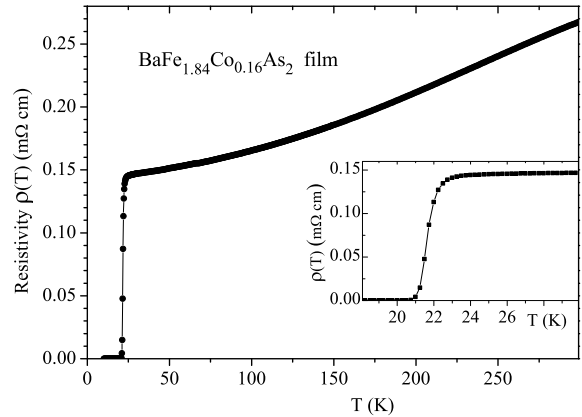}}
 \caption{(Color online) Resistivity $\rho(T)$ of the
BaFe$_{0.84}$Co$_{0.16}$As$_2$ film. In the inset, a zoom of the superconducting
transition. The residual resistivity is
$\rho_0$=$\rho$(25 K)=0.15$\pm$0.01 m$\Omega$cm.}\label{rho}
 \end{figure}

The temperature-dependent resistivity $\rho(T)$ of the investigated film
is reported in Fig.\ref{rho}, showing the onset of the superconducting
transition at T$_c$= 22.5 K with a width of around 2 K. The error bar
in the value of the residual resistivity $\rho_0$ (0.15$\pm$0.01m$\Omega$cm)
is mainly due to uncertainty in the film thickness $t$ estimate (350$\pm$25 nm).

\begin{figure}
\resizebox{0.5\textwidth}{!}{\includegraphics{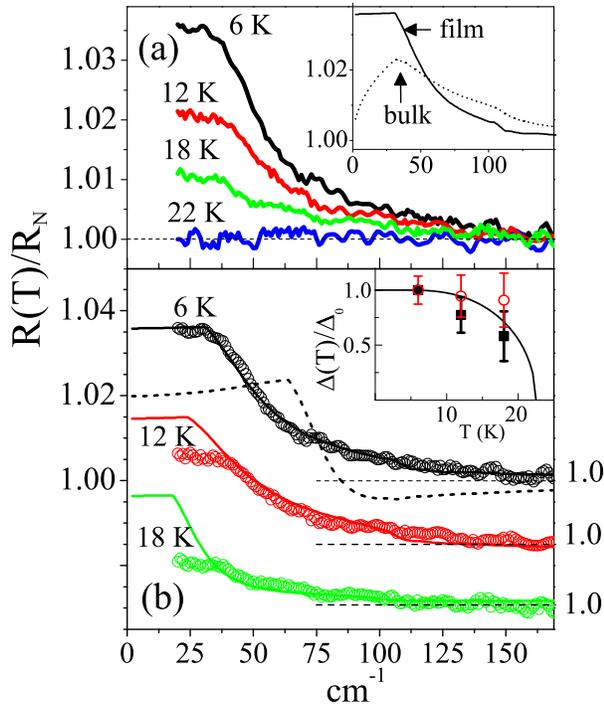}}
\caption{(Color online)(a) $R(T)/R_N$ spectra of the BFCA film in the
THz region. (b) The 6, 12, 18 K spectra conveniently shifted
are compared with the two-band (full line) best fit profiles.
In the 6 K case, also the one-band best fit profile (dotted line) is shown.
In the inset of (a), the $R/R_N$ best-fit model
spectrum of the film is compared with the corresponding spectrum of
a bulk sample (see text).
In the inset of (b), the normalized $\Delta(T)$/$\Delta_0$
values are reported ($\Delta_A$, full squares, $\Delta_B$, open circles),
with $\Delta_0$=$\Delta$(6 K). The full line is the standard BCS
behaviour.} \label{Reflectance_SC}
\end{figure}

We measured the $R(T)/R_N$ spectrum (where $R_N = R$(25 K)) of the
BFCA film in the THz region by employing synchrotron radiation at
the infrared beamline SISSI \cite{SISSI} at the synchrotron Elettra
(Trieste, Italy). We remark that the large size of the film surface,
together with the use of a high-flux synchrotron source, enables
addressing the reflectivity at THz frequencies with superior signal
to noise ratio with respect to measurements on single crystals.
Indeed, our previous $R_S/R_N$ measurements on an ultra-clean
MgB$_2$ film, performed in the THz region by using synchrotron
radiation, provided the first evidence of the effect of the two gaps
in optical measurements \cite{Ort08}, not observed in single
crystals \cite{Kuz07}. The $R(T)/R_N$ measurements were
made by cycling the temperature in the 6-25 K range, without
collecting reference spectra, in order to avoid any variation in the
sample position and orientation, which may yield frequency-dependent
systematic errors in $R(\omega)$. The obtained results are reported
in Fig. \ref{Reflectance_SC}a. We remark
that the high accuracy of the present synchrotron measurements
allows for the detection of small effects of the order of 0.1 \% at
THz frequencies.

For T $\simeq$ T$_c$, the $R(T)/R_N$ curve is flat within
experimental uncertainties, while on decreasing temperature the
$R(T)/R_N$ spectrum increases on decreasing frequency until the
$R_S/R_N$ (with $R_S$=R(6 K)) reaches a plateau at about 30
cm$^{-1}$. At even lower frequencies, $R_S/R_N$ becomes nearly
constant. The frequency dependence of $R_S/R_N$
first suggests the
presence of a superconducting gap $\Delta$ close to 15 cm$^{-1}$.
Indeed, for $\omega\rightarrow0$, the reflectance $R_N$ of a
conducting system tends to 1 for a bulk system, and to a slightly
lower value for a thin film, since the penetration depth becomes
larger than the film thickness. Therefore, since $R_S$ approaches 1
at $\omega=2\Delta$, $R_S/R_N$ exhibits a maximum around $2\Delta$
in the case of a bulk sample, while it remains nearly constant below
$2\Delta$ in the film case\cite{Ort08}.

\section{Analysis and discussion}
For a detailed analysis of the measured spectra, we used the
procedure successfully employed in the MgB$_2$ \cite{Ort08} and
V$_3$Si \cite{v3si} cases. $\tilde{\sigma}_{N}$ is described by the
Drude model, and $\tilde{\sigma}_{S}$ by the Zimmermann model
\cite{Zim91} which generalizes the standard BCS model
\cite{matbar,Tink} to arbitrary temperatures and to systems of
arbitrary purity (i.e., to arbitrary $\gamma$ values).
$\tilde{\sigma}_{S}$ thus depends on the reduced temperature
$t=T/T_c$ and on the parameters $\Omega$, $\gamma$, and $\Delta$.
When two or more bands are assumed to contribute to the film
conductivity, we employed the parallel conductivity model (in the
two-band case \cite{Kuz07,Ort08},
$\tilde{\sigma}=\tilde{\sigma_A}+\tilde{\sigma_B}$), in both the
normal and superconducting states.

From the calculated $\tilde{\sigma}$, by using standard relations
\cite{Burns}, it is then possible to compute the model refractive
index $\tilde{n}=n+ik$ of the film. The reflectance spectrum of the
three-layer system (film-STO-DSO) can then be evaluated by using the
thickness and the $n,k$ values of each layer \cite{Dress}. For STO,
detailed $n$ and $k$ values in the entire infrared region are
reported in the literature \cite{nkSTO}. Since such data are not
available for DSO, we performed absolute reflectance measurements on
a bare DSO substrate over a broad range, thus obtaining the $n$ and
$k$ values relevant to this work. A detailed study of the DSO
optical properties in the far-infrared region will be reported in a
forthcoming paper. We verified that, for any reasonable choice of
the Drude parameters determining $n$ and $k$ of the conducting film,
the intermediate STO layer has no appreciable effect on the THz
reflectance spectrum because of its extremely small thickness.

When only one band is supposed to contribute to the film
conductivity (one-band, 1-b model), the unknown parameters which
determine the $R_S/R_N$ spectrum are $\Omega$, $\gamma$, and $\Delta$.
We thus employed a three-parameters fitting procedure,
and repeated the procedure for different values of the film thickness
($t$=350$\pm$25 nm).
This procedure does not provide an at least satisfactory 
description of the $R_S/R_N$ data independently of the $t$ value, 
as shown by the best-fit curve 
in Fig. \ref{Reflectance_SC}b, obtained for $t$=350 nm 
with $\Omega=3.2$ eV, $\gamma$=800 cm$^{-1}$, and $\Delta$=32 cm$^{-1}$.
The failure of the best fit profile
in reproducing the experimental spectral shape, both above
and below $2\Delta$, indicates the inadequacy of the 1-b model in describing
the optical response of the film at THz frequencies.

In a two-band scenario, the six parameter ($\Omega_i$,$\gamma_i$,
$\Delta_i$, $i=A,B$) fit provides a very good description of the
$R_S/R_N$ spectrum, as shown in Fig. \ref{Reflectance_SC}b. We
remark that only a two-band model can describe the measured
$R_S/R_N$ spectrum, and in particular the smoothly decreasing slope
above $2\Delta$. The fit provides, 
 for $t$=350 nm, plasma frequencies 
$\Omega_A\simeq 1$ eV and
$\Omega_B\simeq$  2 eV, $\gamma_A\simeq$ 200 cm$^{-1}$,
$\gamma_B$ $\simeq$ 2500 cm$^{-1}$, $\Delta_A$ $\simeq$ 15 cm$^{-1}$,
and $\Delta_B$ $\simeq$ 55 cm$^{-1}$.
The $R_S/R_N$ best-fit model spectrum is compared
in the inset of Fig. \ref{Reflectance_SC}b with the model $R_S/R_N$
spectrum evaluated with the same parameters in the bulk case (i.e.,
when $t\rightarrow\infty$). As noted above, the film spectrum is
nearly constant for frequencies lower than the peak frequency of the
bulk spectrum.

In discussing the fit results, we first remark that the Drude
parameters $\Omega_i$ and $\gamma_i$ are affected by rather large
uncertainties (of the order of 20$\%$ 
when the uncertainty in the film 
thickness is considered) since these parameters are
closely inter-related. 
However, the impact of these uncertainties is
reduced in the total plasma frequency
($\Omega_{TOT}=\sqrt{\Omega_A^2+\Omega_B^2}=2.2\pm0.3$ eV) 
and even more in $\sigma_0$=$\sigma_{0A}$+$\sigma_{0B}$
=(7.0$\pm$0.5)10$^3$ $\Omega^{-1}$cm$^{-1}$, corresponding to a residual resistivity 
$\rho_0$=1/$\sigma_0$=0.140$\pm$0.015 m$\Omega$cm, 
in good agreement with the measured value 
$\rho_0$=0.15$\pm$0.01 m$\Omega$cm(see fig.\ref{rho}). 
This result supports the validity of the employed procedure. 
As to the gap values, we can safely pose
$\Delta_A$=15.5$\pm$0.5 cm$^{-1}$ and $\Delta_B$=55$\pm$7 cm$^{-1}$.
Indeed $\Delta_A$ is well determined since, as noted above,
$R_S/R_N$ is nearly constant below $2\Delta_A$, while the $\Delta_B$
value is much more uncertain since this gap does not give a clear
spectral feature. Nonetheless, as discussed above, the B-band
contribution needs to be included in the model in order to obtain a
good description of the experimental $R_S/R_N$ spectral shape.

We then attempted a two-band modelling of the \\
$R(T)/R_N$ spectra
measured at 12 and 18 K by using the 6 K values of the Drude parameters
and using $\Delta_A$ and $\Delta_B$ as free parameters in the fitting procedure.
As shown in Fig. \ref{Reflectance_SC}b, it is not possible to obtain a
satisfactory description of the measured spectra, in particular
at low frequencies. As a consequence, the resulting $\Delta_A$ and $\Delta_B$
values are affected by large uncertainties, as shown in the inset of
Fig. \ref{Reflectance_SC}b where the normalized $\Delta(T)$/$\Delta_0$
values are reported, with $\Delta_0$=$\Delta$(6 K). In the same figure,
the standard BCS $\Delta(T)$/$\Delta_0$ behaviour is reported for comparison.
The poor agreement between the fit and the temperature-dependent data may indicate 
that the two-band model used here still provides an oversimplified description 
of the superconducting gap structure. Thus the possible presence of nodes in the gap 
or of a third gap of very low energy cannot be ruled out.

\begin{figure}
\resizebox{0.5\textwidth}{!}{\includegraphics{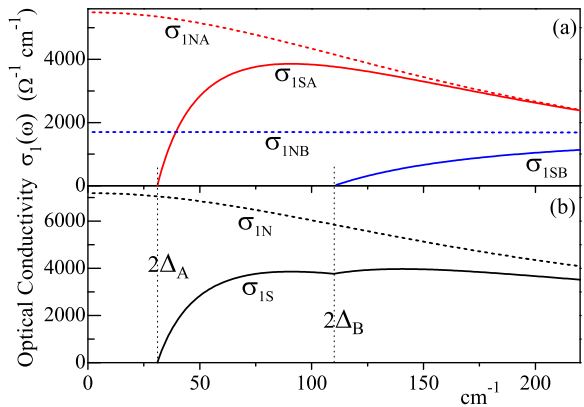}}
\caption{(Color online) Contributions of A and B bands to $\sigma_1$ as
evaluated by using the 2-b best fit parameter values in both the normal
(N) and superconducting (S) state. (b) total $\sigma_{1N}$ and $\sigma_{1S}$.}
\label{opt-cond}
\end{figure}

For a better understanding of the double-gap scenario, we consider
in detail the frequency dependence of the optical conductivity
$\sigma_1$, as evaluated by using the best fit parameter values. The
model $\sigma_{1N}$ and $\sigma_{1S}$ of the A- and B-band are
reported separately in Fig. \ref{opt-cond}a up to 220 cm$^{-1}$,
i.e. in the THz region where the effect of the superconducting gaps
is dominant.
In the normal state, since
$\sigma_{0A}\approx$ 5500 $\Omega^{-1}$ cm$^{-1}$ 
 is roughly three times larger than $\sigma_{0B}$ and
$\gamma_B$ is at least ten times larger than $\gamma_A$,
the spectral shape of the total
$\sigma_{1N}$=$\sigma_{1NA}$+$\sigma_{1NB}$ in the THz region
is dominated by the A-band contribution, as shown in Fig. \ref{opt-cond}b.
 Our results are thus in agreement with recent
findings showing that the normal state conductivity of a
number of 122 systems \cite{wu09}, and in particular of
BaFe$_{2-x}$Co$_{x}$As$_2$ compounds \cite{Nakai10}, is given by
a sharp component superimposed to a much broader component.
In the superconducting state, since $\sigma_{1SA}$ ($\sigma_{1SB}$)
vanishes below 2$\Delta_A$ (2$\Delta_B$), the total
$\sigma_{1S}$=$\sigma_{1SA}$+$\sigma_{1SB}$ vanishes below the
smaller optical gap 2$\Delta_A$, as shown in Fig. \ref{opt-cond}b.
Since, in general, $R_S$ approaches 1 when $\sigma_{1S}$ vanishes,
this explains why only the effect of the smaller gap $\Delta_A$ is
well evident in the $R_S/R_N$ spectrum.

The gap values we obtained are compared in Table \ref{parametri}
with those reported in recent infrared/THz studies of
BaFe$_{2-x}$Co$_{x}$As$_2$ samples. The reflectivity measurements
of van Heumen {\it et al.} on a $x$=0.14 single crystal
\cite{vanh09} provide evidence of a high-energy gap ($\Delta_B$=56
cm$^{-1}$) in very good agreement with our finding, while the
low-energy gap value $\Delta_A$=26.5 cm$^{-1}$ is higher
than our own value (15.5 cm$^{-1}$). Kim {\it et al.} \cite{kim09}
performed reflection and ellipsometry measurements on a $x$=0.13
single-crystal. According to their analysis, three isotropic gaps are 
necessary to obtain a good
description of the experimental data. The low-energy gap $\Delta_A$
value is close to 25 cm$^{-1}$, while the two gaps at higher energy are
around 40 and 78 cm$^{-1}$. Wu {\it et al.} \cite{wu09} made
reflectivity measurements on a $ x$=0.16 crystal and their analysis
provided evidence of a 25 cm$^{-1}$ gap and also suggested the presence
of a second gap at very low energy around 9 cm$^{-1}$.
Gorshunov {\it et al.} \cite{gorsh09}, besides reflectance,
performed transmittance measurements at very low frequencies on a film
\cite{iida09} with $x$=0.2, and found evidence of only one isotropic
gap at 15 cm$^{-1}$, in very good agreement with our $\Delta_A$ value.
 THz conductivity spectroscopy measurements by Nakamura {\it et al.}
\cite{Nakam10} on a $x$=0.2 film showed the presence of a gap at 22.4 cm$^{-1}$. 
The reflectivity measurements of Nakajima {\it et al.} \cite{Nakai10}
on single crystals gave evidence of a single gap at 40 cm$^{-1}$ for
$x$=0.12, around 25 cm$^{-1}$ for $x$=0.16.
Finally, Fisher {\it et al.}
\cite{Fisher10} measured the complex dynamical conductivity of a
$x$=0.2 film and their analysis gave a nodeless gap at 24 cm$^{-1}$
and a nodal gap around 64 cm$^{-1}$.

\begin{table*}
\caption{THz supeconductings gaps in
BaFe$_{2-x}$Co$_{x}$As$_2$ compounds from the present work and
literature \cite{vanh09,kim09,wu09,gorsh09,Nakam10,Nakai10,Fisher10}.}
\label{parametri} \vspace{0.5cm} 
\footnotesize
\begin{tabular}{ccccccccccc}
& Present work  & Ref. \cite{vanh09}  & Ref. \cite{kim09} &
Ref. \cite{wu09}$^a$ & Ref. \cite{gorsh09}  & 
Ref. \cite{Nakam10}$^{b}$ & Ref. \cite{Nakai10}  & Ref. \cite{Nakai10}  & Ref.\cite{Fisher10}  \\
\hline
Co-doping & 0.16 & 0.14 & 0.13 & 0.16 & 0.2 & 0.2 & 0.12 & 0.16  & 0.2\\
 $T_c$ (K)& 22.5 & 23 & 24.5 & 25 & 20 & 19.9 & 25 & 20  & 25\\
N. of gaps &2&2&3&2&1 & 1 & 1 &	1  & 2\\
$\Delta_A$ (cm$^{-1}$) & 15.5 & 26.5 & 25 & 25 & 15 &  22.4 & 40 & 25  & 24  \\
$\Delta_B$ (cm$^{-1}$) & 55 & 56 & 40 & - & - & - & - & - & 64\\
$\Delta_C$ (cm$^{-1}$) & - & - & 78 & - & - & - & -& - & - \\
\end{tabular}
\vspace{0.1cm}

\begin{flushleft}
 $^a$ The authors suggest the presence of a low energy gap
around 9 cm$^{-1}$, below the range in which reliable data could be acquired.

 $^b$ The authors do not exclude an high energy gap since their measurements
are perfomed up to 1 THz (33 cm$^{-1}$).

\end{flushleft}

\end{table*}

\normalsize

Several different measurements therefore indicate a low energy gap
$\Delta_A$ mostrly between 15 and 25 cm$^{-1}$, as reported in Table \ref{parametri}.
In trying to explain this uncertainty, it might be necessary to
consider various effects, 
such as disorder \cite{musch09,Tanatar,Hashimoto09,Carb10} and doping level 
\cite{terashima09} which can differently
affect the various samples, and, in the case of films, the strain imposed
by the substrate \cite{iida09}.
Furthermore, in the case of standard reflectance measurements,
it is important to keep in mind that, in terms of signal to noise ratio,
the quality of THz measurements on a large surface film can be better
than that of measurements on crystals, especially when the wavelength
approaches the crystal size.
The difference among the observed
$\Delta_A$ values might then be explained, but more bothersome is
the disagreement in the number and values of the high-energy gaps,
as summarized in Table \ref{parametri}. 

\section{Conclusions}
We measured the THz reflectance properties of a high quality
epitaxial film of BaFe$_{1.84}$Co$_{0.16}$As$_2$ with T$_c$=22.5 K
 by measuring the $R(T)/R_N$ spectrum, i.e. the reflectivity 
ratio between the superconducting state (for 6$\leq$T$\leq$ 22 K)
and the 25 K (normal state) reflectance.
The measured $R_S/R_N$ spectrum (with $R_S=R(6 K)$ 
first suggests the presence of a superconducting gap $\Delta_A$ 
close to 15 cm$^{-1}$.
We find that only a two-band model, with a gap opening in
both bands, can describe the measured $R_S/R_N$ spectrum, and in particular
its smoothly decreasing slope.
 The low-energy $\Delta_A$ gap ($\Delta_A$=15.5$\pm$0.5 cm$^{-1}$) and
the high-energy gap $\Delta_B$ ($\Delta_B$=55$\pm$7 cm$^{-1}$) correspond
to 2$\Delta/kT_c$ values close to 2 and 7, respectively.
The $R_S/R_N$ data are thus compatible with a 
two-band scenario, with a nodeless isotropic gap opening in both bands. 
Nevertheless, the poor agreement between the model and experimental
$R(T)/R_N$ at 12 and 18 K can be interpreted as an indication for 
ungapped states at low energies. Their presence may be attributed either 
to nodes or to the presence of gaps at energies so low  as to get 
thermally excited carriers at relatively modest temperatures.

Despite some differences, the magnitude of the low-energy optical gap
recently found by different groups substantially agree with
our $\Delta_A$ value, but there is a large disparity on the reported
number of gaps and on their values. In order to resolve this
important issue, THz transmission measurements on
BaFe$_{2-x}$Co$_{x}$As$_2$ films, to be performed in the entire THz
region where gap effects are expected, could be of great help in
settling this debate. Indeed, in a multigap superconductor,
transmittance measurements are more sensitive to the features
related to the high-energy gap(s), while reflectance is more
dependent on the low-energy one, as we have shown in the V$_3$Si
case \cite{v3si}.

\section*{Acknowledgements}
Work at NHMFL was supported under NSF Cooperative Agreement DMR-0084173,
by the State of Florida, and by AFOSR grant FA9550-06-1-0474.
Work at the University of Wisconsin was supported by the US Department
of Energy, Division of Materials Science,
under Award No. DE-FG02-06ER463.
The authors thank M.S. Rzchowski for helpful discussion.
Work at the University of Rome was partially funded by CARIPLO Foundation
(Project 2009-2540 ''Chemical Control and Doping Effects in Pnictide
High-temperature Superconductors''), at the University of Genova
by the "Italian Foreign Affairs
Ministry (MAE) - General Direction for the Cultural Promotion".

\end{document}